\documentstyle[12pt,epsf]{article}
\def\be{\begin{equation}}
\def\ee{\end{equation}}
\def\xt{{\tilde x}}
\def\Pt{{\widetilde P}}
\def\ffrac#1#2{\textstyle{#1\over#2}\displaystyle}
\gdef\journal#1, #2, #3, 1#4#5#6{{\sl #1~}{\bf #2}, #3, 1#4#5#6}

\begin{document}
\baselineskip=24pt
\pagestyle{empty}
\vspace{-1mm}
\begin{flushright}
JC-98-06\\
cond-mat/9812416
\end{flushright}
\vspace{5mm}
\begin{center}
{\bf \LARGE Critical Exponents near a Random Fractal Boundary\\}
\vspace{8mm}
{\bf \large John Cardy\\}
\vspace{2mm}
{Department of Physics\\
Theoretical Physics\\1 Keble Road\\Oxford OX1 3NP, UK\\
\& All Souls College, Oxford\\}
\end{center}
\vspace{6mm}
\begin{abstract}
The critical behaviour of correlation functions near a boundary is
modified from that in the bulk. When the boundary is smooth this is
known to be characterised by the surface scaling dimension $\xt$.
We consider the case when the boundary is a random fractal, specifically a
self-avoiding walk or the frontier of a Brownian walk, in two
dimensions, and show that the boundary scaling behaviour of the 
correlation function is characterised by a set of multifractal boundary
exponents, given exactly by conformal invariance arguments to be 
$\lambda_n=\frac1{48}(\sqrt{1+24n\xt}+11)(\sqrt{1+24n\xt}-1)$.
This result may be interpreted in terms of a scale-dependent
distribution of opening angles $\alpha$ of the fractal boundary: on short
distance scales these are sharply peaked around $\alpha=\pi/3$.
Similar arguments give the multifractal exponents for the case of coupling to a
quenched random bulk geometry.

\end{abstract}
\newpage

\pagestyle{plain}
\setcounter{page}{1}
\setcounter{equation}{0}

The subject of boundary critical behaviour \cite{reviews} is by now well
understood, particularly in two dimensions \cite{JCpapers}. The
two-point correlation function $\langle\phi(r)\phi(R)\rangle$ of a
scaling operator $\phi$, which behaves in the bulk at large distances
at the critical point as $|r-R|^{-2x}$, where $x$ is the bulk scaling
dimension of $\phi$, is modified when one of the points (say $r$) is
close to the boundary to the form 
\be
\label{0}
\langle\phi(r)\phi(R)\rangle \sim |r|^{-x}|R|^{-x}|R/r|^{-\xt}, 
\ee
where $\xt$ is the corresponding boundary scaling dimension, and 
the angular dependence has been suppressed for clarity. In two
dimensions, the role played by $\xt$ is emphasised by making 
the conformal mapping $z\to\ln z$ of the upper half plane to a strip of
width $\pi$: in that geometry the correlation function decays exponentially
along the strip with an inverse correlation length equal to $\xt$
\cite{JCamp}.

Eq.~\ref{0} refers to the case when the boundary is smooth (at least on scales
$\ll r$) and it is an interesting to ask whether these results are
modified when the boundary is a fractal on these scales. The example
of an edge (or corner in two dimensions) on the boundary was
analysed some time ago \cite{JCedge} and it was shown that new edge
scaling dimensions arise which depend \em continuously \em on the
opening angle $\alpha$. In two dimensions \cite{JCcorner} this dependence
is given by conformal invariance arguments by the simple form
$x(\alpha)=\pi\xt/\alpha$. This suggests that close to a fractal
boundary, which may be thought of as presenting a distribution of
opening angles (which perhaps also depends on the scale at which it is
probed), an even more complicated behaviour should obtain.

In the case of a random fractal, one also expects to see behaviour
characteristic of correlation functions in a quenched random
environment. That is, they may exhibit \em multiscaling\em, which means
that the average of their $n$th power does not scale in the same way as
the $n$th power of their average. In this letter, we consider two cases
where this problem is exactly solvable using conformal invariance
methods in two dimensions, namely when the fractal boundary is a
self-avoiding walk, and when it is the frontier (exterior boundary)
 of a Brownian
(ordinary) random walk. In fact, both cases turn out to give identical
results. Our methods are a simple generalisation of arguments due
to Lawler and Werner \cite{LW}, who have derived exact relations between
multifractal exponents
corresponding to self-intersection properties of Brownian walks in two
dimensions. This corresponds to the special case when $\phi$ is a free scalar
field satisfying Laplace's equation. This physically interesting 
example, and its relation to the exponents of star polymers,
was in fact discussed some time ago in $4-\epsilon$ dimensions
by Cates and Witten \cite{CW}.
The results of Lawler and Werner have recently been given an elegant
interpretation and derivation by Duplantier \cite{Dup98}
in the context of coupling the system to a randomly
fluctuating metric.

\begin{figure}
\centerline{
\epsfxsize=5in
\epsfbox{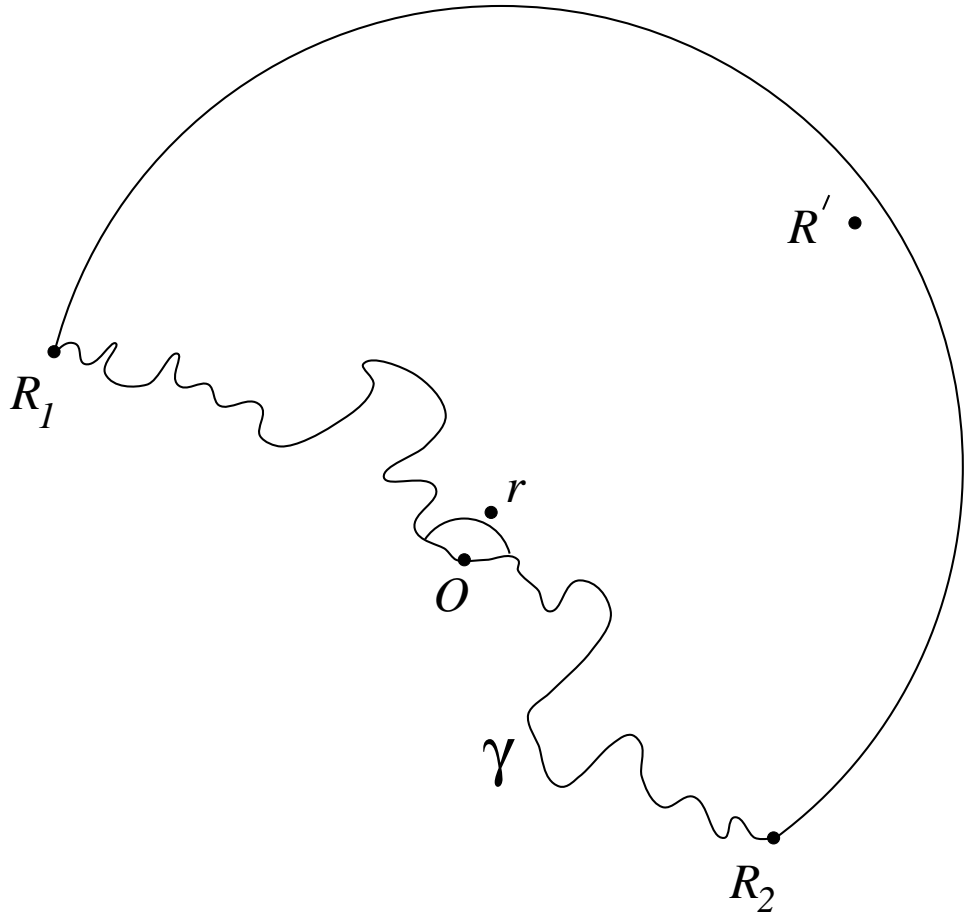}}
\caption{Geometry in which the simply connected region $R_1OR_2$ is
bounded by a self-avoiding walk $\gamma$ and an arc of the circle $\rho=R$. This
region contains a critical system of which the correlation function of
local operators at points $r$ and $R'$ is of interest. This
region, excluding the disc $\rho<|r|$, is conformally mapped into
a long strip of width $\pi$ and length $\ell_{\rm eff}(\gamma,r/R)$.}
\label{fig1}
\end{figure}

Consider for definiteness self-avoiding walks $\gamma$ 
which are constrained to pass through the origin $O$.
In order to be able to apply conformal invariance arguments, we work in
the fixed fugacity ensemble, in which each walk of length $L$ is
counted with a weight $y^L$, at the critical point where $y^{-1}=\mu$,
the lattice-dependent connective constant. The properties of the measure
on walks
on distance scales much larger than the lattice spacing are then supposed
to be conformally invariant. Denote the radial coordinate by $\rho$.
We want to focus on those walks which have a typical linear size $R$,
and for which the origin $O$ is a typical interior point. Without
loss of generality for computing scaling dimensions, we may then
restrict the walks $\gamma$ to have the form of a pair of mutually
avoiding self-avoiding walks, starting from $O$ and ending on the
circle $\rho=R$. In the same spirit we may take these points to be the
first intersections of the walks with this circle. The region bounded by
$\gamma$ and an arc of the circle $\rho=R$ is thus simply connected.
In this region we consider a critical system (for example an Ising
model) with a suitable conformally invariant boundary condition \cite{JCpapers}
on $\gamma$ and on $\rho=R$ (for example, that the spins are free).
Consider the correlation function $\langle\phi(r)\phi(R')\rangle$,
where, without loss of generality for computing scaling dimensions
\cite{CW} we can choose $|R'|\sim R$, and we are interested in the limit
where $r\ll R$. The geometry is illustrated in Fig.~\ref{fig1}. 

Obviously this correlation function depends on $\gamma$, but we may hope to
be able to compute suitable averages of this quantity over realisations
of $\gamma$. By analogy with the case of a smooth
boundary, we expect that
\be
\label{1}
\overline{\langle\phi(r)\phi(R')\rangle^n}
\sim |r|^{-nx}|R|^{-nx}|R/r|^{-\lambda_n},
\ee
where the overline means an average over all the allowed realisations
of $\gamma$. 
Note that if the average were over smooth
boundaries only, we would expect $\lambda_n=n\xt$, with only the
prefactor modified by the averaging.

We now make a conformal transformation which maps the fractal
boundary into a smooth one. It is convenient to exclude the disc
$\rho<r$. This leaves the simply connected region bounded by two
segments of $\gamma$ and arcs of the circles $\rho=r$ and $\rho=R$. 
By the Riemann theorem, the interior of this region may be mapped
conformally by an analytic function $z\to f(\gamma,|r|,R;z)$
onto the interior of a strip of width $\pi$, but with a length
which is not simply $\ell\sim\ln(R/r)$, but which will also depend
on $\gamma$. Let us denote this by $\ell_{\rm eff}(\gamma,\ell)$. 
The correlation function $\langle\phi(r)\phi(R')\rangle$ will be related
by this conformal mapping to one between operators 
located near the ends of this strip.
Taking the $n$th power and averaging we see by
comparison with (\ref{1}) that
\be
\label{3}
e^{-\lambda_n\ell}\sim \overline{e^{-n\xt\ell_{\rm eff}(\gamma,\ell)}}
\ee

At this point, we need further information about averages of quantities
of the form $e^{-p\ell_{\rm eff}}$, for arbitrary $p$. 
In particular, let us consider the case
where $n=1$, and $\phi(r)$ is the $M$-leg operator for $M$ mutually
self-avoiding walks. It is convenient to generalise slightly and take
$\phi(R')$ to be a product of distinct single leg operators 
corresponding to the walks all ending on the arc of the circle $\rho=R$.
The correlation function then gives 
the number of such walks which all begin at $r$
and end at a distance $\sim R$, and, in the critical fugacity
ensemble, scales in the bulk like $|r-R|^{-x_M-Mx_1}$, and near a smooth
boundary according to (\ref{0}) with a boundary scaling dimension $\xt_M$.
Coulomb gas arguments \cite{DupSal} lead to the conjectures
$x_M=\frac3{16}M^2-\frac1{12}$ and $\xt_M=\frac18M(3M+2)$, which have
been confirmed by numerical work and various other known exact
information. If we now imagine taking an $M$-leg star polymer near the
fractal boundary $\gamma$, which itself is a 2-leg star
polymer, and performing the same average over the realisations of $\gamma$,
the result will be an $(M+2)$-leg star polymer in the bulk.
We conclude that
\be
\label{4}
e^{-(x_{M+2}-x_2)\ell}\sim \overline{e^{-\xt_M\ell_{\rm eff}(\gamma,\ell)}}\quad,
\ee
where the factor $e^{x_2\ell}$ arises from the normalisation of the
probability distribution of $\gamma$. 
Note that although this is initially defined only for $M$ a non-negative
integer, it may be continued to other real values for which the average
exists. Comparing with (\ref{3}) we may therefore choose $M$ such that
$\xt_M=n\xt$, solve for $M$ using the exact conjecture for $\xt_M$ given
above, and substitute this into the exact form for $x_{M+2}$. After some
simple algebra, this gives the result for $\lambda_n$ quoted in the
abstract.

A similar argument may be made when $\gamma$ is the exterior boundary of
a Brownian walk. In this case the measure is rigorously known to be 
conformally invariant. As in the example considered by Lawler
and Werner \cite{LW}, one may now invoke the exact
conjectures made by Duplantier and Kwon \cite{DupKwon}
for the relevant dimensions of
$M$ mutually avoiding \em ordinary \em random walks. However, the final result
is identical, indicating that there is strong element of
universality between these two fractals, not only with respect to their
fractal dimensions \cite{Mandelbrot}. 

The nontrivial dependence of $\lambda_n$ shows that correlation
functions near the boundary have a broad distribution of values. In
particular, their average value, which scales as $(r/R)^{\lambda_1}$, 
may be quite
different from a typical value. As argued, for example, in
Ref.~\cite{Ludwig}, the \em typical \em dependence should be of the form
$(r/R)^{\lambda'}$, where $\lambda'=d\lambda_n/dn|_{n=0}=3\xt$.
Interesting enough this is the behaviour which would obtain in wedge
of interior angle $\alpha=\pi/3$. This idea may be made more explicit by
interpreting (\ref{3}) in terms of an average over a scale-dependent
distribution $P(\alpha,\ell)$ of interior opening angles: 
\be
\int_0^{2\pi}d\alpha\,P(\alpha,\ell)\,e^{-(\pi/\alpha)n\xt\ell}
\sim e^{-\lambda_n\ell}
\ee
Requiring that this be valid for all positive real $n$ determines the form
of $P$. First we see that the behaviour of $\lambda_n\sim n\xt/2$ as
$n\to\infty$ at fixed $\ell$ implies that the effective angle in this
regime is $\alpha\sim2\pi$. This is in agreement with a general argument
of Cates and Witten \cite{CW}. If we set $\omega\equiv\frac\pi\alpha-\frac12$
and $u\equiv n\xt$, and define $\Pt(\omega,\ell)d\omega=
P(\alpha,\ell)d\alpha$ the above equation simplifies to
\be
\int_0^\infty d\omega\,\Pt(\omega,\ell)\,e^{-\omega u\ell}
\sim \exp\left(-(5\ell/24)(\sqrt{1+24u}-1)\right)\quad.
\ee
Making the ansatz $P(\omega,\ell)\sim e^{5\ell/24}\,e^{-\ell(a\omega
+b/\omega)}$ and using steepest descent then leads to the solution
\be
\Pt(\omega,\ell)\sim \exp\left(-\frac\ell{24}\left(\sqrt\omega-\frac5{2\sqrt
\omega}\right)^2\right)\quad,
\ee
valid for large $\ell$ (i.e. $r/R\ll1$) and where we have suppressed
more slowly varying prefactors. Note that this result is independent of $\xt$,
consistent with it being an intrinsic property of the fractal\footnote{However,
it should be noted that the method of averaging used here, which sums
over all realisations of $\gamma$ passing through a given point $O$ at
given distance $r$ from a fixed point, tends to
emphasise those parts of the fractal for which the opening angle is
small.}.
It shows that the effective opening angle has a broad distribution
which, however, becomes more and more strongly peaked around the typical
value $\omega=\frac52$ ($\alpha=\frac\pi3$) as $r/R\to0$, with a width of
order $(\ln(R/r))^{-1/2}$. 

A similar calculation may be carried out when the point $O$ is the
root of an $N$-leg star polymer, by replacing $x_{M+2}-x_2$ in 
(\ref{4}) by $x_{M+N}-x_N$. The case $N=1$ gives the end
multifractal exponents, which, as first pointed out by Cates and Witten,
are different form those which arise when $O$ is a typical interior
point.

In the case considered by Cates and Witten \cite{CW},
where $\phi$ is a massless scalar field (with bulk
scaling dimension $x=0$) satisfying Dirichlet conditions $\phi=0$ on the
boundary, the appropriate boundary scaling operator is
$\partial_\perp\phi$ with dimension $\xt=1$. In that case we find that
$\lambda(1)=\frac23=2-D$, where $D=\frac43$ is the fractal dimension of
the boundary. This is a consequence of the fact that $\phi$ satisfies
Laplace's equation, equivalent to the conservation of particle flux in
the Brownian interpretation \cite{CW}. Note that we were lead to this
unique result from making simple assumptions (rigorously grounded in the
Brownian case) about the conformal invariance of the measure on $\gamma$.
This suggests that all such curves which, with probability one, bound a
simply connected region when viewed on macroscopic distance scales
will fall into this universality class and, in particular, will have 
$D=\frac43$.

One of the most interesting features of our main result is that the
$\lambda_n$ are generally, even for integer $n$, irrational (but
algebraic) numbers. This is not in disagreement with any established
results, since even if the bulk critical theory is unitary, 
there is no reflection positivity in the presence of a fractal boundary
and so the theorem of Friedan, Qiu and Shenker \cite{FQS} is evaded. 
However, most
examples of exactly calculable critical exponents in two dimensions have,
even in nonunitary cases, led to rational values.
Recently Duplantier \cite{Dup98} has given an interesting interpretation
of these type of results in the case of a general mixture of
Brownian and self-avoiding walks, by considering
the effects of coupling the system to a fluctuating background metric
(quantum gravity).
He was able to argue that, just as the scaling dimensions of
overlapping objects in flat space should be added to obtain that of the
composite, when they are coupled to quantum gravity their dressed
scaling dimensions are additive if they \em avoid \em each other.
In this way, by going back and forth between flat space and quantum
gravity, and using the relation between ordinary and dressed scaling
dimensions first obtained by Knizhnik, Polyakov and Zamolodchikov (KPZ)
\cite{KPZ},
specialised to the case $c=0$ appropriate to Brownian walks, he was
able not only to recover the results of Lawler and Werner \cite{LW}, but
also to derive the earlier conjecture of Duplantier and Kwon
\cite{DupKwon}. Thus, for this example, his methods are more powerful
than the simple arguments we have used above, since we found it necessary
to invoke the conjectured values for the $M$-leg scaling dimensions.
Since our main result is a simple generalisation to the case when
$\xt\not=1$, one would expect that similar quantum gravity methods might
apply. However, our result is supposed to be valid for
a bulk theory with arbitrary central charge $c$, and it is therefore not
clear why the KPZ relation with $c=0$ should appear in this more general
case.

We have given an exact formula for the multiscaling boundary exponents
of an arbitrary conformally invariant two-dimensional critical system
close to a random fractal boundary. This is the first example when such
a multifractal spectrum with a non-trivial analytic structure has been
found exactly. The basic method was to realise that this kind of
geometric quenched disorder may be gauged away by making a suitable
conformal transformation, at the cost of modifying the moduli (in this
case $\ell=\ln(R/r)$.) The effective distribution of $\ell_{\rm eff}$
is then probed by replacing the critical system by one with $c=0$
(in our case, self-avoiding walks) for which the partition function is
unity and therefore the quenched average of a correlation function is
the same as its annealed average, which is more simply dealt with.

Similar ideas may be applied to a critical system coupled to 
\em bulk \em quenched disorder in the form of a random metric
(which may be realised as the continuum limit of a randomly connected
lattice.) In this case we may consider an annulus of inner and outer
radii $r$ and $R$ respectively, which in the case of a flat metric may
be mapped conformally by the transformation $z\to\ln z$ to a flat metric
on a cylinder of perimeter $2\pi$ and length $\ell=\ln(R/r)$.
The inverse correlation length along this cylinder is then equal to the bulk
scaling dimension $x$ \cite{JCamp}. An arbitrary metric $g$ on the
annulus is also conformally equivalent to the cylinder with a flat
metric but with a length $\ell_{\rm eff}(g,\ell)$. In analogy with
(\ref{3}) the multifractal bulk exponents $\lambda_n^b$, which govern
the decay of the quenched average of the $n$th power of the bulk
correlation function, are given by
\be
e^{-\lambda_n^b\ell}\sim \overline{e^{-nx\ell_{\rm eff}(g,\ell)}}
\ee
We now consider the special case when the critical system case has
$c=0$. In that case the quenched and annealed averages are identical,
and the respective scaling dimensions $X_0$ and $X$ of an operator 
in flat space and when coupled to a fluctuating metric are related by
the $c=0$ version of the KPZ relation \cite{KPZ} 
$X_0=\frac13X(1+X)$. Thus if we now set $X_0=nx$ and solve for $X$, this
will yield $\lambda_n^b$. The result is
\be
\lambda_n^b=\ffrac12\left(\sqrt{1+12nx}-1\right)\quad.
\ee
This result for the scaling dimensions when coupled to a quenched random
lattice, was derived for the case $n=1$ by Baillie, Hawick and Johnston
\cite{BHJ}, but in fact these
correlation functions exhibit multiscaling, with
a whole spectrum of such exponents. Note that in this case
the typical decay of a correlation function is determined by
$(\lambda^b)'=3x$. 

The author thanks G.~Lawler and B.~Duplantier for explaining their
ideas, and the Fields Institute, Toronto, where this work was started,
for its hospitality.  This research was partly supported by the
Engineering and Physical Sciences Research Council under Grant GR/J78327.

After this work was completed preprints by Duplantier
\cite{Dup99} and Aizenman, Duplantier and Aharony \cite{ADA} appeared
in which,
among other things, it is argued that the fractal dimension of the
accessible perimeter of a percolation cluster has fractal dimension 
$D=\frac43$, consistent with the general arguments advanced above.

\end{document}